\documentclass[aps]{revtex4}
\usepackage{amssymb}
\usepackage[dvips]{graphicx}
\usepackage{amsmath}
\usepackage{graphicx}
\usepackage{makeidx}
\usepackage{subfigure}

\begin{document}

\title{Stabilizing single- and two-color vortex beams in quadratic media by
a trapping potential}
\author{Hidetsugu Sakaguchi$^{1}$ and Boris A. Malomed$^{2}$}
\address{$^{1}$Department of Applied Science for Electronics and Materials,\\
Interdisciplinary Graduate School of Engineering Sciences,\\
Kyushu University, Kasuga, Fukuoka 816-8580, Japan\\
$^{2}$Department of Physical Electronics, School of Electrical Engineering,\\
Faculty of Engineering, Tel Aviv University, Tel Aviv 69978, Israel}

\begin{abstract}
We consider two-dimensional (2D) localized modes in the
second-harmonic-generating ($\chi ^{(2)}$) system with the
harmonic-oscillator (HO) trapping potential. In addition to its
realization in optics, the system describes the mean-field dynamics
of mixed atomic-molecular Bose-Einstein condensates (BECs). The
existence and stability of various modes is determined by their
total power, $N$, topological charge, $m/2$ [$m$ is the intrinsic
vorticity of the second-harmonic (SH) field], and $\chi ^{(2)}$
mismatch, $q$. The analysis is carried out in a numerical form and,
in parallel, by means of the variational approximation (VA), which
produces results that agree well with numerical findings. Below a
certain power threshold, $N\leq N_{c}^{(m)}(q)$, all trapped modes
are of the \textit{single-color} type, represented by the SH
component only, while the fundamental-frequency (FF) one is absent.
In contrast with the usual situation, where such modes are always
unstable, we demonstrate\ that they are \emph{stable}, for $m=0,1,2$
(the mode with $m=1$ may be formally considered as a
\textit{semi-vortex} with topological charge $m/2=1/2$), at $N\leq
N_{c}^{(m)}(q)$, and unstable above this threshold. On the other
hand, $N_{c}^{(m)}(q)\equiv 0$ at $q\geq q_{\max }$ (in our
notation, $q_{\max }=1$), hence the single-color modes are unstable
in the latter case. At $N=N_{c}^{(m)}$, the modes with $m=0$ and
$m=2$ undergo a pitchfork bifurcation, which gives rise to two-color
states, which remain completely stable for $m=0$. The two-color
vortices with $m=2$ (topological charge $1$) have an upper stability
border, $N=N_{c2}(q)$. Above the border, they exhibit periodic
splittings and recombinations, while keeping their vorticity. The
semi-vortex does not bifurcate; at $N=N_{c}^{(m=1)}$, it exhibits
quasi-chaotic oscillations and a rotating ``groove" resembling a
screw-edge dislocation induced by the semi-integer vorticity.
\end{abstract}

\maketitle

OSIS numbers: 190.6135; 190.3100; 190.4410; 020.475 \bigskip

\section{Introduction}

It is commonly known that the quadratic, alias $\chi ^{(2)}$, nonlinearity,
plays a fundamental role in nonlinear optics, helping, in particular, to
create various species of solitons \cite{review0}-\cite{review}, \cite{KA}.
The use of the $\chi ^{(2)}$ nonlinearity is crucially important for the
making of 2D and 3D solitons, because, on the contrary to Kerr ($\chi ^{(3)}$%
) nonlinearity, the quadratic interaction between the FF and SH fields does
not give rise to the collapse \cite{Rubenchik}, which is a severe problem
for the stability of 2D and 3D solitons in Kerr media \cite{review}. Thanks
to this circumstance, the first $\chi ^{(2)}$ solitons were created as
stable (2+1)-dimensional beams propagating in an SH-generating crystal \cite%
{first-exper}. Further, the absence of the collapse instability in the 3D
setting suggests a possibility of creating fully localized ``light bullets"
\cite{Drummond}. In the experiment, 3D solitons have not been observed yet,
the best result being a spatiotemporal soliton self-trapped in the
longitudinal and one transverse directions, due to the interplay of the
diffraction, group-velocity dispersion, and $\chi ^{(2)}$ nonlinearity,
while the confinement in the other transverse direction was provided by the
waveguiding structure \cite{Wise1,Wise2}.

Another natural possibility in the (2+1)D setting is the creation of
vortical solitary beams, with the ``hollow" in the middle. In these
modes, self-trapped SH and FF fields carry intrinsic vorticities $m$
and $m/2$, respectively. The modes are classified as vortex solitons
with topological charge $m/2$; solitons with odd values of $m$ are
not possible, as the intrinsic vorticity of the FF component, $m/2$,
cannot take half-integer values, although vortices with a
half-integer optical angular momentum can be created, in the form of
mixed screw-edge dislocations, by passing the holding beam through a
spiral-phase plate displaced off the beam's axis \cite{half}. Unlike
their fundamental counterparts with $m=0$, the vortex solitons in
the free space are always unstable against azimuthal perturbations,
which split them into sets of separating segments. This instability
was predicted theoretically \cite{splitting1}-\cite{splitting5} and
demonstrated in the experiment \cite{splitting-exp}. The same
instability was also predicted in the framework of the so-called
Type-II (three-wave) $\chi ^{(2)}$ system, which includes two
distinct components of the FF field
\cite{splitting-3W,splitting-3W2}.

Solitons with embedded vorticity are also known as solutions to the 2D
nonlinear Schr\"{o}dinger (NLS) equation with the self-focusing cubic term
\cite{Kruglov}, and they too are subject to the azimuthal instability, which
is actually stronger than the collapse-induced instability \cite{review}.
The 2D self-focusing NLS equation models not only the light transmission in
bulk media with the Kerr nonlinearity, but also [in the form of the
Gross-Pitaevskii (GP) equation] the mean-field dynamics of BECs in ultracold
gases with attractive inter-atomic interactions, shaped as ``pancakes" by
the confining potential \cite{BEC}. A solution to the instability problem
was elaborated in the latter context: both fundamental solitons and solitary
vortices with topological charge $1$ can be stabilized by isotropic HO
(harmonic-oscillator) trapping potentials. As shown in detail in a number of
theoretical works \cite{cubic-in-trap1}-\cite{cubic-in-trap7}, the HO
potential stabilizes the fundamental solitons in the entire region of their
existence, while vortex solitons are stabilized, in terms of their norm (the
counterpart of the total power of spatial optical solitons), in $\simeq 33\%$
of their existence region, and in an adjacent region of width $\simeq 10\%$
vortices exist in the form of periodically splitting and recombing modes,
which keep their vorticity \cite{cubic-in-trap5}.

The effective 2D trapping potential can be also realized in optical
waveguides, in the form of the respective profile of the transverse
modulation of the local refractive index \cite{KA}. This circumstance
suggests a natural possibility for the stabilization of (2+1)D vortex
solitons in the $\chi ^{(2)}$ medium by means of the radial HO potential,
which is the main subject of the present work. A feasible approach to the
making of the optical medium combining a nearly-parabolic profile of the
refractive index and $\chi ^{(2)}$ nonlinearity is the use of a 2D photonic
crystal, which can be readily designed to emulate the required index
profile, while the nonlinearity is provided by the poled material (liquid
\cite{Du} or solid \cite{Luan}) filling the voids. As shown below, the
effective radial potential provides for sufficiently strong localization of
the trapped modes, therefore the exact parabolic shape of the radial profile
is not crucially important. The analysis can be readily adjusted to other
profiles, if necessary.

The model, based on the system of coupled equations for the FF and SH
fields, is introduced in Section II. It is relevant to mention that
essentially the same system of GP equations for the atomic and molecular
mean-field wave functions describes the BEC in the atomic-molecular mixture
\cite{BEC0}-\cite{BEC4}. Accordingly, the predicted mechanism of the
stabilization of two-component vortex solitons trapped in the HO potential
can also be realized in the BEC mixture.

Solutions for the trapped 2D modes and their stability against perturbations
are considered in Sections III and IV. First, we address states which, in
the unperturbed form, contain only the SH field, while the FF component
vanishes. Such \textit{single-color} solutions of the $\chi ^{(2)}$ system
are known in other contexts, but they are usually subject to the parametric
instability against small perturbations in the FF component. Our first
result is that the \emph{trapped} single-color modes, both fundamental and
vortical ones, have \emph{a finite stability domain}. In particular, the
single-color modes with $m=1$, which (formally) look as \textit{semi-vortices%
}\emph{\ }with topological charge $1/2$, are also found, and they are stable
too in a finite parameter area. The single-color states become unstable at a
particular critical value of the total power (alias norm); however, the
critical norm vanishes if the mismatch parameter, $q$, is too large, \textit{%
viz}., at $q>1$ in the notation adopted below). As concerns the fundamental
modes ($m=0$) and those with topological charge $1$ ($m=2$), at exactly the
same critical point they undergo a pitchfork bifurcation, which gives rise
to two-color states. The stability of the fundamental two-color mode is
obvious, while a nontrivial result is finding stability borders for the
trapped two-color vortex soliton. Above the instability border, it develop
periodic oscillations, keeping its vorticity and featuring periodic
splittings and recoveries. As concerns the single-color semi-vortex, it
exhibits a different behavior, developing persistent quasi-random
oscillations above the stability border, mixing the first and zeroth angular
harmonics in both the SH and FF components.

The results concerning the shape and stability of the modes of all the
above-mentioned types (single-color ones with $m=0,1,2$, and two-color
complexes with $m=0$ and $2$) are obtained, in parallel, by means of
numerical methods and in an analytical form, based on the variational
approximation (VA). In almost all the cases, the VA demonstrates very good
accuracy in comparison with numerical results.

\section{The model}

The model for the SH generation in 2D is based on the usual scaled equations
\cite{review0}-\cite{review2} for the FF and SH field amplitudes, $u$ and $v$%
,
\begin{eqnarray}
i\frac{\partial u}{\partial z}+\frac{1}{2}\left( \frac{\partial ^{2}}{%
\partial x^{2}}+\frac{\partial ^{2}}{\partial y^{2}}\right) u+u^{\ast
}v-U(x,y)u &=&0,  \notag \\
2i\frac{\partial v}{\partial z}+\frac{1}{2}\left( \frac{\partial ^{2}}{%
\partial x^{2}}+\frac{\partial ^{2}}{\partial y^{2}}\right) v-qv+\frac{1}{2}%
u^{2}-4U(x,y)v &=&0,  \label{eqs}
\end{eqnarray}%
where $z$ is the propagation distance, $x$ and $y$ are the transverse
coordinates, the asterisk stands for the complex conjugate, $q$ is the real
mismatch coefficient, and, as said above, the axisymmetric modulation of the
refractive index is modeled by the isotropic HO potential, $U(x,y)=\left(
\Omega ^{2}/2\right) \left( x^{2}+y^{2}\right) $. By means of an obvious
rescaling, we fix $\Omega =1/2$, while $q$ remains a free parameter.

The relation between the potential terms in the equation for the FF and SH
fields implies that the same refractive index acts on both fields, which is
a realistic assumption for materials of which the above-mentioned pair of
the photonic crystal and voids-filling stuff can be fabricated. If the weak
index dispersion is taken into regards, it will produce only a small
perturbation in the system.

The 2D system of scaled GP equations for the atomic-molecular mixture
corresponds to Eqs. (\ref{eqs}) with $z$ replaced by time $t$, and $q$
accounting for a difference of the chemical potential between the atomic and
molecular components, $u$ and $v$. It is relevant to mention that a similar
model with a spatially periodic (lattice) potential $U\left( x,y\right) $
was considered in Ref. \cite{chi2-in-lattice}, where it was demonstrated
that the lattice can readily stabilize two-color solitary vortices, although
with an anisotropic shape.

Equations (\ref{eqs}) are derived from the corresponding action, $S=\int Ldz$%
, with Lagrangian
\begin{eqnarray}
L &=&\int \int \left\{ \left[ iu_{z}u^{\ast }+2iv_{z}v^{\ast }-\frac{1}{2}%
\left( |u_{x}|^{2}+|u_{y}|^{2}+|v_{x}|^{2}+|v_{y}|^{2}\right) \right] \right.
\notag \\
&&\left. -U(r)\left( |u|^{2}+4|v|^{2}\right) -q|v|^{2}+\frac{1}{2}\left(
u^{2}v^{\ast }+u^{2\ast }v\right) \right\} dxdy.  \label{L}
\end{eqnarray}%
Stationary modes can be characterized by their total power (norm), $N=\int
\int \left( |u|^{2}+4|v|^{2}\right) dxdy$.

\section{Stability of single-color beams}

\subsection{The beams with topological charge $0$ and $1$ ($m=0$ and $m=2$)}

For the single-color modes with $u=0$, the SH field satisfies the 2D linear
Schr\"{o}dinger equation with the isotropic HO potential,
\begin{equation}
2i\frac{\partial v}{\partial z}+\frac{1}{2}\left( \frac{\partial ^{2}}{%
\partial x^{2}}+\frac{\partial ^{2}}{\partial y^{2}}\right) v-qv-4U(x,y)v=0.
\label{linear}
\end{equation}%
Stationary solutions to Eq.~(\ref{linear}) are commonly known from quantum
mechanics. In polar coordinates $\left( r,\theta \right) $, they are
\begin{equation}
v=v_{m0}\exp \left( i\left( -\mu z+m\theta \right) \right) r^{m}\exp \left(
-r^{2}/2\right) ,  \label{singleSH}
\end{equation}%
with arbitrary amplitude $v_{m0}$ (it is defined to be real), integer
orbital quantum number $m$ (as said above, it corresponds to the beam's
topological charge $m/2$), and eigenvalue $\mu =\left( m+1+q\right) /2$
(recall $\Omega =1/2$ is fixed). The total power of this solution is
\begin{equation}
N_{m}=4\pi m!v_{m0}^{2}.  \label{N(m=0)}
\end{equation}

\begin{figure}[tbp]
\begin{center}
\includegraphics[height=5.5cm]{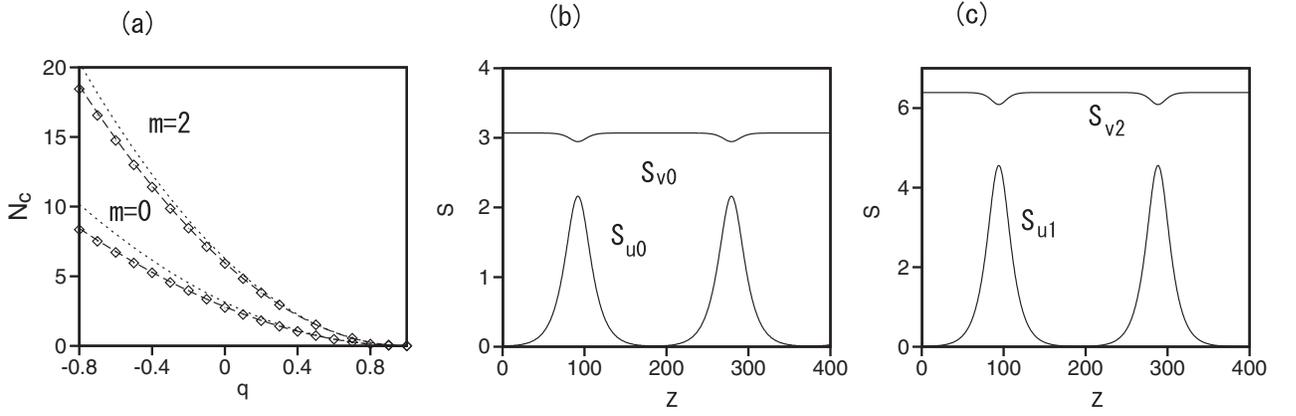}
\end{center}
\caption{(a) The critical power for the onset of the parametric instability
of the single-color beams with topological charges $0$ ($m=0$) and $1$ ($m=2$%
), the beams being unstable at $N>N_{c}$. Chains of rhombuses and dashed and
dotted curves show, respectively, numerical results and the prediction of
the variational approximation produced by Eqs. (\protect\ref{Nc(m=0)}), (%
\protect\ref{Nc(m=2)}), and (\protect\ref{0}). Examples of periodic
oscillations of perturbed solutions at $N>N_{c}$ are shown for $m=0,q=0$,
and $N=3$ in (b), and for $m=2$, $q=0$, and $N=6.5$ in (c). Amplitudes $%
S_{u0,v0}(z)$ and $S_{u1,v2}$ are defined by Eqs. (\protect\ref{S00}) and (%
\protect\ref{S12}), respectively.}
\label{fig1}
\end{figure}

Solutions (\ref{singleSH}) are obviously stable within the framework of
linear equation (\ref{linear}), the issue being to find a threshold, $%
N=N_{c}^{(m)}(q)$, at which the parametric instability against infinitesimal
perturbations in the FF field sets in, due to the nonlinearity in Eq. (\ref%
{eqs}) for the FF field. As shown in Fig. \ref{fig1}(a), the threshold was
identified from systematic simulations of the perturbed evolution of the
single-color beams within the framework of the full system of Eqs. (\ref{eqs}%
), for $m=0$ and $2$.

The stability was also investigated in an analytical form by means of the
VA. For $m=0$, the ansatz for the perturbed solution is taken as
\begin{equation}
u=u_{0}(z)\exp \left( -\alpha r^{2}\right) ,~v=v_{0}(z)\exp \left(
-r^{2}/2\right) ,  \label{ans(m=0)}
\end{equation}%
where amplitudes $u_{0}(z)$ and $v_{0}(z)$ are treated as variational
parameters, while $\alpha $ is a free constant, which is used below as a
variational parameter too, but in a different sense. The substitution of
ansatz (\ref{ans(m=0)}) into Lagrangian (\ref{L}) leads, after
straightforward calculations, to the following Euler-Lagrange equations for
variables $u_{0}(z)$ and $v_{0}(z)$:
\begin{eqnarray}
i\frac{du_{0}}{dz} &=&\left( \alpha +\frac{1}{16\alpha }\right) u_{0}-\frac{%
4\alpha }{4\alpha +1}u_{0}^{\ast }v_{0},  \label{u0} \\
i\frac{dv_{0}}{dz} &=&\left( \frac{1}{2}+\frac{q}{2}\right) v_{0}-\frac{2}{%
4\alpha +1}u_{0}^{2}\text{.}  \label{v0}
\end{eqnarray}

The unperturbed single-color solution to Eqs. (\ref{u0}), (\ref{v0}), $%
u_{0}=0$, $v_{0}(z)=v_{00}\exp \left( -\left( i/2\right) \left( 1+q\right)
z\right) $, coincides with exact solution (\ref{singleSH}) (with $m=0$).
Further, to investigate the onset of the parametric instability against the
excitation of the infinitesimal FF perturbation, we linearize Eq. (\ref{u0})
and substitute $u_{0}(z)\equiv u_{0}^{\prime }(z)\exp \left( -\left(
i/4\right) \left( 1+q\right) z\right) $, which leads to the following
equation for the perturbation's amplitude:
\begin{equation}
i\frac{du_{0}^{\prime }}{dz}=\left( \alpha +\frac{1}{16\alpha }-\frac{1}{4}-%
\frac{q}{4}\right) u_{0}^{\prime }-\frac{4\alpha }{4\alpha +1}v_{00}\left(
u_{0}^{\prime }\right) ^{\ast }.  \label{u0'}
\end{equation}%
An elementary consideration of Eq. (\ref{u0'}) demonstrates that it gives
rise to the parametric instability at $\left[ \alpha +1/(16\alpha )-1/4-q/4%
\right] ^{2}<16\alpha ^{2}/(4\alpha +1)^{2}v_{00}^{2}$, hence the critical
value of total power (\ref{N(m=0)}) is
\begin{equation}
N_{c}^{(m=0)}=4\pi \left( \alpha +\frac{1}{16\alpha }-\frac{1}{4}-\frac{q}{4}%
\right) ^{2}\left( \frac{4\alpha +1}{4\alpha }\right) ^{2}.  \label{Nc(m=0)}
\end{equation}

A similar stability analysis can be performed for the vortical single-color
beam (\ref{singleSH}) with $m=2$. In this case, the natural ansatz for the
FF perturbation is taken in the form of the first angular harmonic, i.e.,
\begin{equation}
u=u_{1}^{\prime }(z)r\exp \left[ -\alpha r^{2}+i\theta -\left( i/4\right)
\left( 3+q\right) z\right] ,  \label{u1}
\end{equation}%
which, after straightforward calculations, leads to the following evolution
equation [cf. Eq. (\ref{u0'})]:
\begin{equation}
i\frac{du_{1}^{\prime }}{dz}=\left( 2\alpha +\frac{1}{8\alpha }-\frac{3}{4}-%
\frac{q}{4}\right) u_{1}^{\prime }-\frac{64\alpha ^{2}}{(4\alpha +1)^{3}}%
v_{20}\left( u_{1}^{\prime }\right) ^{\ast }.  \label{u1'}
\end{equation}%
Taking into regard expression (\ref{N(m=0)}), Eq. (\ref{u1'}) yields the
critical power,
\begin{equation}
N_{c}^{(m=2)}=\frac{\pi (4\alpha +1)^{6}}{512\alpha ^{4}}\left( 2\alpha +%
\frac{1}{8\alpha }-\frac{3}{4}-\frac{q}{4}\right) ^{2}.  \label{Nc(m=2)}
\end{equation}

The final analytical prediction for the destabilization thresholds is
obtained by the minimization of each expression, (\ref{Nc(m=0)}) and (\ref%
{Nc(m=2)}) with respect to the variation of free parameter $\alpha $, for
given $q$. In particular, an explicit result of the minimization is that, in
both cases of $m=0$ and $m=2$,
\begin{equation}
\begin{array}{c}
N_{c}^{(m=0,2)}\equiv 0~~\mathrm{at}~~q\geq 1, \\
N_{c}^{(m=0,2)}\approx \pi \left[ 1+(m/2)\right] \left( 1-q\right) ^{2}~~%
\mathrm{at}~~0<1-q\ll 1.%
\end{array}
\label{0}
\end{equation}%
The so obtained critical values of $N_{c}$ are shown in Fig.~\ref{fig1}(a)
by dashed lines, which approximate the numerical results very accurately.
Note that Eq. (\ref{0}) explains the vanishing of $N_{c}$ at $q>1$, which is
obvious in Fig. \ref{fig1}(a).

As concerns the realization of the instability in the optical waveguide, it
is relevant to note that its length is finite (and usually not very large)
in a real experiment. For this reason, the observable instability threshold
may be shifted to somewhat larger values of $N$, as a very small instability
growth rate would not be able to manifest itself on a relatively short
propagation distance.

Above the instability threshold, i.e., at $N>N_{c}^{(m)}$, simulations of
both Eqs. (\ref{eqs}) and VA equations, (\ref{u0}) and (\ref{v0}),
demonstrate oscillatory behavior of perturbed solutions. An example is
displayed in Fig. \ref{fig1}(b), at $m=0$, $q=0$, and $N=3$, for variables
which, essentially, measure the amplitudes of the zeroth angular harmonic in
the FF and SH fields:
\begin{equation}
S_{u0,v0}(z)\equiv \left\vert \int \int \left\{ u\left( x,y\right) ,v\left(
x,y\right) \right\} dxdy\right\vert .  \label{S00}
\end{equation}%
It is observed in the figure that the unperturbed state with the zero FF
amplitude is periodically recovered. A similar example for $m=2$, $q=0$, and
$N=6.5$ is displayed in Fig. \ref{fig1}(c), for the integral amplitudes of
the first and second angular harmonics in the FF and SH fields,
respectively:
\begin{equation}
S_{u1,v2}(z)\equiv \left\vert \int \int \left\{ u\left( x,y\right)
e^{-i\theta },v\left( x,y\right) e^{-2i\theta }\right\} dxdy\right\vert .
\label{S12}
\end{equation}

\subsection{Half-vortices: the beams with topological charge $1/2$ ($m=1$)}

A noteworthy peculiarity of the single-color beams (\ref{singleSH}) is that
they may have $m=1$, which formally correspond to the topological charge $%
m/2\equiv 1/2$. Of course, this is only possible due to the fact that the FF
field is absent in the stationary solution; nevertheless, it is shown in
what follows below that the half-integer charge of the unperturbed solution
essentially affects the perturbed evolution of the single-color vortex above
the instability threshold.

To test the stability of these \textit{half-vortices }within the framework
of the VA, a natural ansatz for the FF perturbation is defined as a
combination of the zeroth and first angular harmonics: $u=u_{0}^{\prime
}\exp \left( -\alpha r^{2}-i\gamma z\right) +u_{1}^{\prime }r\exp \left(
-\alpha ^{\prime }r^{2}+i\theta -i\gamma ^{\prime }z\right) $, with $\gamma $
and $\gamma ^{\prime }$ related by the necessary matching condition,
\begin{equation}
\gamma +\gamma ^{\prime }=1+q/2.  \label{match}
\end{equation}%
The VA gives rise to linear coupled equations for perturbation amplitudes $%
u_{0}^{\prime }$ and $u_{1}^{\prime }$ [cf. Eqs. (\ref{u0'}) and (\ref{u1'}%
)]:
\begin{eqnarray}
\gamma u_{0}^{\prime } &=&\left( \alpha +\frac{1}{16\alpha }\right)
u_{0}^{\prime }-\frac{4\alpha }{2\left( \alpha +\alpha ^{\prime }\right) +1}%
v_{10}\left( u_{1}^{\prime }\right) ^{\ast },  \notag \\
\gamma ^{\prime }u_{1}^{\prime } &=&\left( 2\alpha ^{\prime }+\frac{1}{%
8\alpha ^{\prime }}\right) u_{1}^{\prime }-\frac{16\alpha ^{\prime 2}}{\left[
2\left( \alpha +\alpha ^{\prime }\right) +1\right] ^{2}}v_{10}\left(
u_{0}^{\prime }\right) ^{\ast }.  \label{u01'}
\end{eqnarray}

\begin{figure}[tbp]
\begin{center}
\includegraphics[height=5.4cm]{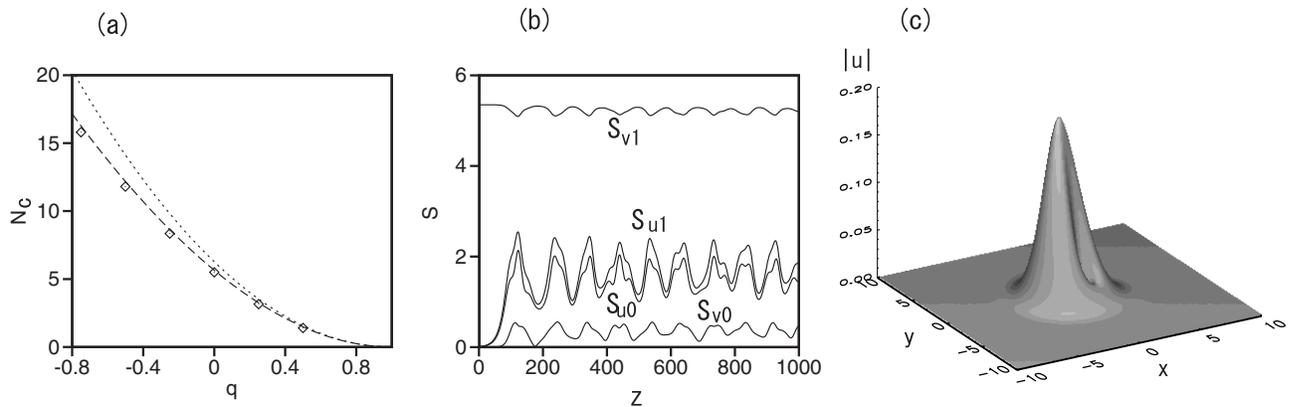}
\end{center}
\caption{(a) The critical power for the onset of the parametric instability
of the single-color \textit{semi-vortex} with topological charge $1/2$ ($m=1$%
). Chains of rhombuses and the dashed and dotted curves show, respectively,
numerical results and the prediction of the variational approximation
produced by Eqs. (\protect\ref{Nc(m=1)}) and (\protect\ref{0'}). (b) An
example of quasi-chaotic oscillations of an unstable perturbed solution is
shown for $m=1,~q=0$, and $N=5.8$. (c) The 3D profile of the FF component of
the same solution at $z=1000$. }
\label{fig2}
\end{figure}

A straightforward analysis of Eqs.~(\ref{u01'}) yields the following
resolvability condition:
\begin{equation}
\frac{128\alpha \alpha ^{\prime 2}v_{10}^{2}}{\left[ 2\left( \alpha +\alpha
^{\prime }\right) +1\right] ^{4}}=\left( \alpha +\frac{1}{16\alpha }-\gamma
\right) \left( 2\alpha ^{\prime }+\frac{1}{8\alpha ^{\prime }}-\gamma
^{\prime }\right) .  \label{alpha}
\end{equation}%
Substituting relation (\ref{match}) in Eq.~(\ref{alpha}), one obtains a
quadratic equation for $\gamma $. The parametric instability sets in at a
critical value of the total power, $N=N_{c}^{(m=1)}$, when the discriminant
of the quadratic equation vanishes, which leads to the following result:
\begin{gather}
N_{c}^{(m=1)}=\frac{\pi (\alpha +\alpha ^{\prime }+1/2)^{4}}{2\alpha \alpha
^{\prime 2}}  \notag \\
\times \left[ \left( \alpha ^{\prime }+\frac{1}{16\alpha ^{\prime }}-\frac{1%
}{2}-\frac{q}{4}-\frac{\alpha }{2}-\frac{1}{32\alpha }\right) ^{2}\right.
\notag \\
\left. +\left( \alpha +\frac{1}{16\alpha }\right) \left( 2\alpha ^{\prime }+%
\frac{1}{8\alpha ^{\prime }}-1-\frac{q}{2}\right) \right] .  \label{Nc(m=1)}
\end{gather}%
The expression on the right-hand side should be minimized with respect to
the variation of $\alpha $ and $\alpha ^{\prime }$. The so generated
critical (dashed) curve $N_{c}^{(m=1)}(q)$ is displayed in Fig.~\ref{fig2}%
(a) along with the results produced by direct simulations of Eqs.~(\ref{eqs}%
). Like in the situation displayed for $m=0$ and $m=2$ in Fig.~\ref{fig1}%
(a), the variational prediction is very close to its numerical counterpart.
Further, also similar to what was done above, an explicit result can be
obtained from Eq. (\ref{Nc(m=1)}) in the following form [cf. Eq. (\ref{0})]:
\begin{equation}
\begin{array}{c}
N_{c}^{(m=1)}\equiv 0~~\mathrm{at}~~q\geq 1, \\
N_{c}^{(m=1)}\approx 2\pi \left( 1-q\right) ^{2}~~\mathrm{at}~~0<1-q\ll 1,%
\end{array}
\label{0'}
\end{equation}%
see the dotted line in Fig. \ref{fig2}(a). Incidentally, it coincides with
Eq. (\ref{0}) with $m=2$.

At $N>N_{c}^{(m=1)}$, the perturbed evolution demonstrates persistent
quasi-chaotic oscillations of the FF and SH amplitudes, on the contrary to
perfectly periodic oscillations in the cases of $m=0$ and $2$, cf. Fig. \ref%
{fig1}(b,c). A typical example is displayed in Fig. 2(b) in terms of the
integral amplitudes defined as per Eqs. (\ref{S00}) and (\ref{S12}), for $q=0
$ and $N=5.8$. In particular, it is observed that the instability generates
the zeroth angular harmonic in the SH field, represented by amplitude $S_{v0}
$, which was absent in stationary solution (\ref{singleSH}), that included
solely the second angular harmonic. Figure \ref{fig2}(c) displays, for the
same solution, a 3D plot of $\left\vert u(x,y)\right\vert $ at $z=1000$. The
groove structure observed in the plot is explained by the fact that $%
|u\left( x,y\right) |=\left\vert u_{0}^{\prime }(z)\exp (-i\gamma z-\alpha
r^{2})+u_{1}^{\prime }(z)r\exp (-i\gamma ^{\prime }z-\alpha ^{\prime
}r^{2}+i\theta )\right\vert $ becomes small near a certain value of angle $%
\theta $. The dynamics of the groove is further illustrated in Fig. \ref%
{fig3} by a set of three contour plots of $|u\left( x,y\right) |$, plotted
at $z=1000,~1002$ and $1004$. It is observed that the groove rotates
counter-clockwise. This feature resembles the above-mentioned mixed
screw-edge dislocation carried by the beam with the half-integer vorticity
\cite{half}, although the amplitude does not vanish in the groove, and the
inspection of the respective phase field does not feature a clear jump by $%
\pi $, which may be explained by the fact that the quasi-chaotic dynamics
stirs the phase structure.
\begin{figure}[tbp]
\begin{center}
\includegraphics[height=5.4cm]{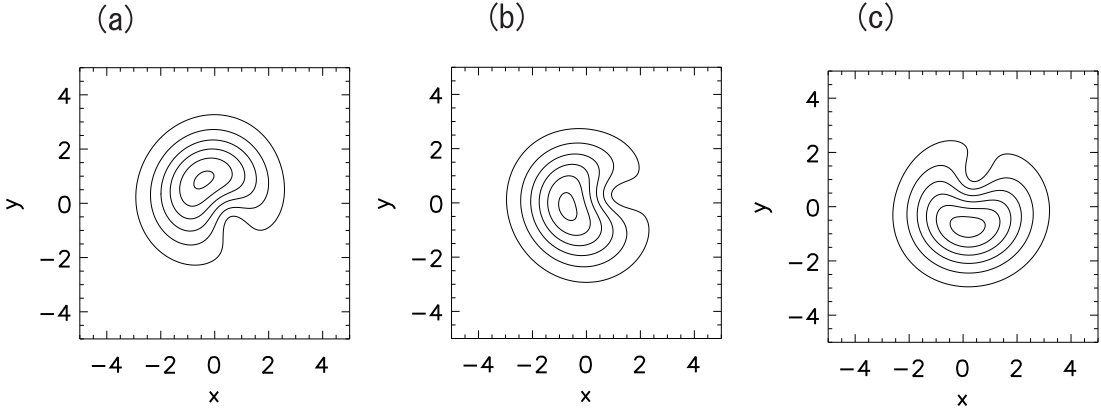}
\end{center}
\caption{Contour plots of the FF component of the same solution which is
displayed in Figs. \protect\ref{fig2}(b,c), at $z=1000$ (a), $z=1002$ (b),
and $z=1004$ (c).}
\label{fig3}
\end{figure}

\section{The stability of two-color beams}

While the single-color beams, with $u=0$, are unstable at $N>N_{c}^{(m)}$,
in precisely the same region there appear two-color modes for $m=0$ and $m=2$
, built of the FF ($u\neq 0$) and SH fields. In other words, $N_{c}^{(m)}$
is not only the stability border for the trapped single-color modes, but
also the existence threshold for their two-color counterparts. It is
relevant to stress that, as seen from Fig. \ref{fig1} and Eqs. (\ref{0}) and
(\ref{0'}), the threshold vanishes at $q\geq 1$.

It is easy to construct the two-color solutions for $m=0$, in the form of $%
u\left( x,y,z\right) =e^{-i\mu z}u_{0}(r),~v\left( x,y,z\right) =e^{-2i\mu
z}v_{0}(r)$. These solutions exist at $N>N_{c}^{(m=0)}$ as completely stable
modes, which is natural, as 2D fundamental solitons in the $\chi ^{(2)}$
system are stable even in the absence of the trapping potential \cite%
{review1,review2}.

The transition from the solutions with $u=0$ to $u\neq 0$, with the
simultaneous destabilization of the single-color state, is a pitchfork
bifurcation, which happens with the increase of $N$, as is shown in Fig. \ref%
{fig4} for $m=2$ (the pitchfork must give rise to two mutually symmetric
modes at $N>N_{c}$, which corresponds to the obvious fact that any solution
with $u\neq 0$ has its counterpart with the same $v$ and $u\rightarrow -u$).
Of course, such an transition does not occur for the single-color
half-vortices with $m=1$, as in that case the FF field, $u\left(
x,y,z\right) \neq 0$, would carry intrinsic vorticity $m/2=1/2$, which is
impossible.
\begin{figure}[tbp]
\begin{center}
\includegraphics[height=6.cm]{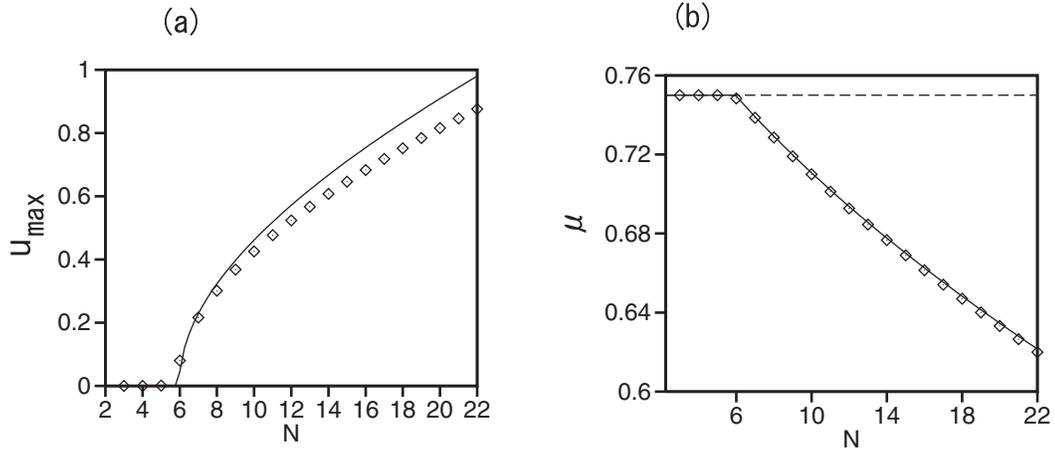}
\end{center}
\caption{Amplitude $u_{\max }$ of the FF field (a) and propagation constant
(b) of the solitary-vortex beams with $m=2$ at $q=0$, of both the single-
and two-color types, at $N<N_{c}^{(m=2)}$ and $N>N_{c}^{(m=2)}$,
respectively. The pitchfork bifurcation occurs at $N=N_{c}^{(m=2)}$. Chains
of rhombuses and the dashed curve show, respectively, numerical results and
the prediction of the variational approximation based on ansatz (\protect\ref%
{ans(m=2)}).}
\label{fig4}
\end{figure}

For the vortex with $m=2$, the solution extended past the bifurcation point
can be looked for as per the ansatz
\begin{equation}
u\left( x,y,z\right) =u_{10}r\exp \left( -i\mu z+i\theta -\alpha
r^{2}\right) ,\ v\left( x,y,z\right) =v_{20}r^{2}\exp \left( -2i\mu
z+2i\theta -\beta r^{2}\right) ,  \label{ans(m=2)}
\end{equation}%
whose total power is [cf. Eq. (\ref{N(m=0)})]%
\begin{equation}
N=\pi \left[ \left( 2\alpha \right) ^{-2}u_{10}^{2}+\beta ^{-3}v_{20}^{2}%
\right] ,  \label{N}
\end{equation}%
with variational parameters $u_{10}$, $v_{20}$ and $\alpha $, $\beta $.
Figure \ref{fig4} shows the amplitude of $u_{10}$ and propagation constant $%
\mu $ for the vortex beams with $m=2$ at $q=0$, as predicted by the VA on
the basis of this ansatz, and as found from numerical solutions. The dashed
line in Fig. \ref{fig4}(b) represents the constant value, $\mu =3/4+q/2$,
for the single-color vortex with $m=2$, which is unstable at $%
N>N_{c}^{(m=2)} $. The VA is very accurate close to the bifurcation point,
showing a discrepancy which slowly increases at large $N$.

A crucially important issue is the stability of the vortex beams at $%
N>N_{c}^{(m=2)}$, as all such states are unstable against azimuthal
splitting in the free space \cite{splitting1}-\cite{splitting-exp}. In the
analytical form, the stability can be explored by mean of the nonstationary
version of the VA, based on the following ansatz, which adds perturbations
containing spatial harmonics with numbers $-1$ and $3$ in the FF field, and
zeroth and fourth harmonics in the SH component, to the stationary solution
taken as per Eq. (\ref{ans(m=2)}) (such perturbations induce the splitting
instability of the solitary vortex in the free space):
\begin{eqnarray}
u\left( x,y,z\right) &=&u_{1}(z)r\exp \left( i\theta -\alpha r^{2}\right)
+u_{-1}(z)r\exp \left( -i\theta -\alpha _{-1}r^{2}\right) +u_{3}(z)r^{3}\exp
\left( 3i\theta -\alpha _{3}r^{2}\right) ,  \notag \\
v\left( x,y,z\right) &=&v_{2}(z)r^{2}\exp \left( 2i\theta -\beta
r^{2}\right) +v_{0}(z)\exp \left( -\beta _{0}r^{2}\right) +v_{4}(z)r^{4}\exp
\left( 4i\theta -\beta _{4}r^{2}\right) .  \label{uv}
\end{eqnarray}%
Here $\alpha $ and $\beta $ are the same as found from the stationary
version of the VA based on ansatz (\ref{ans(m=2)}). Straightforward
calculations lead to the following linearized evolution equations for the
perturbation amplitudes:
\begin{eqnarray}
i\frac{du_{-1}}{dz} &=&\left( 2\alpha _{-1}+\frac{1}{8\alpha _{-1}}\right)
u_{-1}-\frac{4\alpha _{-1}u_{1}^{\ast }v_{0}}{(\alpha +\alpha _{-1}+\beta
_{0})^{2}}-\frac{24\alpha _{-1}^{2}u_{3}^{\ast }v_{2}}{(\alpha _{-1}+\alpha
_{3}+\beta )^{4}},  \notag \\
i\frac{du_{3}}{dz} &=&\left( 4\alpha _{3}+\frac{1}{4\alpha _{3}}\right)
u_{3}-\frac{16\alpha _{3}^{4}u_{-1}^{\ast }v_{2}}{(\alpha _{-1}+\alpha
_{3}+\beta )^{4}}-\frac{64\alpha _{3}^{4}u_{1}^{\ast }v_{4}}{(\alpha +\alpha
_{3}+\beta _{4})^{4}},  \notag \\
i\frac{dv_{0}}{dz} &=&\left[ \beta _{0}+\frac{1}{2\left( 4\beta
_{0}+q\right) }\right] v_{0}-\frac{\beta _{0}u_{1}u_{-1}}{(\alpha +\alpha
_{-1}+\beta _{0})^{2}},  \notag \\
i\frac{dv_{4}}{dz} &=&\left[ \frac{5\beta _{4}}{2}+\frac{1}{2\left( 4\beta
_{4}+q\right) }\right] v_{4}-\frac{16\beta _{4}^{5}u_{1}u_{3}}{(\alpha
+\alpha _{3}+\beta _{4})^{5}}.  \label{evolution}
\end{eqnarray}%
Solutions to Eqs. (\ref{evolution}) are looked for as $%
u_{-1}=u_{-1}^{(0)}e^{-i\gamma _{-1}z},u_{3}=u_{3}^{(0)}e^{-i\gamma
_{3}z},v_{0}=v_{0}^{(0)}e^{-i\gamma _{0}z}$, $v_{4}=v_{4}^{(0)}e^{-i\gamma
_{4}z}$, with propagation constants subject to the matching conditions,
which ensue from the substitution of expressions (\ref{ans(m=2)}) into Eqs. (%
\ref{evolution}):
\begin{equation}
\gamma _{0}=\mu +\gamma _{-1},\gamma _{3}=2\mu -\gamma _{-1},\gamma
_{4}=3\mu -\gamma _{-1}.  \label{gamma}
\end{equation}%
Finally, the eigenvalue problem for $\gamma _{-1}$ amounts to the following
equation, in which $\gamma _{0}$, $\gamma _{3}$, and $\gamma _{4}$ should be
substituted as per Eq. (\ref{gamma}):
\begin{gather}
\frac{6\alpha _{-1}^{2}\alpha _{3}^{4}\left( 8v_{20}\right) ^{2}}{(\alpha
_{-1}+\alpha _{3}+\beta )^{8}}=\left[ 2\alpha _{-1}+\frac{1}{8\alpha _{-1}}%
-\gamma _{-1}-\frac{32\alpha _{-1}^{2}\beta _{0}u_{10}^{2}}{(\alpha +\alpha
_{-1}+\beta _{0})^{4}(4\beta _{0}+\beta _{0}^{-1}+4q-8\gamma _{0})}\right]
\notag \\
\times \left[ 4\alpha _{3}+\frac{1}{4\alpha _{3}}-\gamma _{3}-\frac{2\alpha
_{3}^{4}\beta _{4}^{5}\left( 64u_{10}\right) ^{2}}{(\alpha +\alpha
_{3}+\beta _{4})^{10}(20\beta _{4}+5\beta _{4}^{-1}+4q-8\gamma _{4})}\right]
.  \label{eigen}
\end{gather}

\begin{figure}[tbp]
\begin{center}
\includegraphics[height=6.cm]{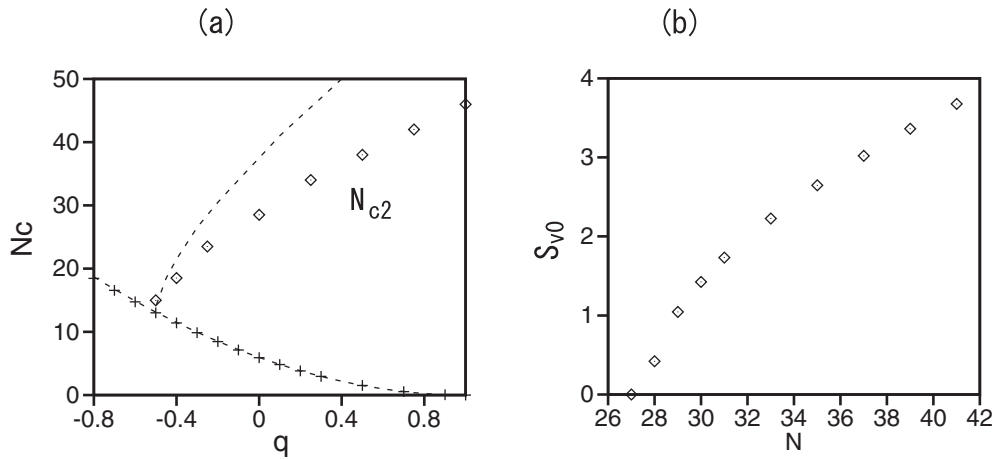}
\end{center}
\caption{(a) The stability area of the two-color vortices with topological
charge $1$ ($m=2$), in the plane of the mismatch ($q$) and total power ($N$%
), is $N_{c}<N<N_{c2}$. The numerically found stability borders and their
counterparts predicted by the variational approximation are displayed by
chains of symbols and dashed lines, respectively [in fact, $N_{c}(q)$ is the
same border as one labeled by $m=2$ in Fig. \protect\ref{fig1}(a)]. (b) The
peak value of the integral amplitude of the zeroth angular harmonic of the
SH field, $S_{v0}(z)$ [defined as per Eq. (\protect\ref{S00})], as found
from the simulations of the oscillatory solutions at $N>N_{2c}\approx 27.5$
for $q=0$.}
\label{fig5}
\end{figure}

The critical value of the total power [see Eq. (\ref{N})] at the onset of
the instability of the trapped vortex (\ref{ans(m=2)}), $N=N_{c2}$, is
identified as the one at which eigenvalue $\gamma _{-1}$, found from Eq. (%
\ref{eigen}), becomes complex. Then, it should be minimized by varying the
set of parameters $\alpha _{-1},\alpha _{3},\beta _{0}$ and $\beta _{3}$,
cf. Eqs. (\ref{Nc(m=0)}), (\ref{Nc(m=2)}), and (\ref{Nc(m=1)}). Actually,
this procedure is too cumbersome in its full form, but we have found that a
nearly-minimum value of $N_{c2}$ corresponds to $\alpha _{-1}=\alpha
_{3}=1.2\alpha $ and $\beta _{0}=\beta _{3}=0.5$. The result is reported in
Fig. \ref{fig5}(a), which displays $N_{c2}$ versus $q$, along with the
previously found critical value, $N_{c}$, which is the border between the
stable single-color vortices and emerging stable two-color ones. As seen
from the figure, the two-color vortex is completely unstable at $q<q_{\min
}\approx -0.5$, and has an expanding stability area at $q>q_{\min }$. The
increase of $N_{c2}$ with mismatch $q$ is explained by the fact that large $%
q $ corresponds to the cascading limit \cite{review0}-\cite{review2}, in
which the $\chi ^{(2)}$ nonlinearity is transformed into the self-focusing
cubic interaction, with a decreasing effective cubic coefficient, $\chi
^{(3)}\sim 1/q$, and the vortices trapped by the HO potential in the
respective \emph{weakly nonlinear} cubic medium have a large stability area
\cite{cubic-in-trap1}-\cite{cubic-in-trap7}. The accuracy of the VA
prediction for $N_{c2}$ is essentially lower than it was for $N_{c}$,
because the variational \textit{ans\"{a}tze} (\ref{ans(m=2)}) and (\ref{uv})
cannot approximate the complex structure of the respective modes accurately
enough, and by the above-mentioned fact that the full minimization procedure
is too cumbersome in this case [in particular, we expect that a more
thorough procedure would make the VA-predicted values of $N_{c2}$ somewhat
smaller than those plotted in Fig. \ref{fig5}(a), thus reducing the
discrepancy with the numerically found stability border]. Nevertheless, the
VA predicts the value of $q=q_{\min }$ almost exactly, and the general shape
of the stability boundary, $N=N_{c2}(q)$, is predicted correctly too.

\begin{figure}[tbp]
\begin{center}
\includegraphics[height=6.cm]{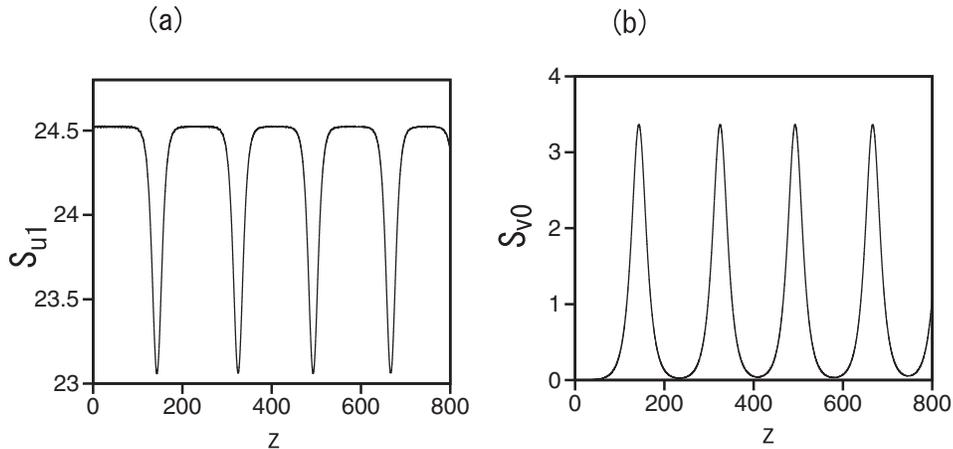}
\end{center}
\caption{The periodic evolution of the integrally defined amplitudes of the
first angular harmonic in the FF field (a), and the zeroth harmonic in the
SH field (b).}
\label{fig6}
\end{figure}

At $N>N_{c2}$, the simulations demonstrate that the instability transforms
the stationary two-color vortex into a persistent oscillatory one, which
keeps the vortical component, mixing it with the zero-vorticity one. To
illustrate this finding, the peak values (largest over the period of the
oscillations) of the integrally defined amplitude of the zeroth angular
component in the SH field, $S_{v0}(z)$ [see Eq. (\ref{S00})], is displayed
as a function of $N$ in Fig. \ref{fig5}(b) for $q=0$. The persistent
oscillatory dynamics developed by the unstable two-color vortices at $%
N>N_{c2}$ is further illustrated in Fig. \ref{fig6} by the evolution of the
integrated amplitudes $S_{u1}(z)\ $and $S_{v0}(z)$ [defined as per Eqs. (\ref%
{S12}) and (\ref{S00})], produced by the direct simulations for $q=0$ and $%
N=39$, cf. Fig. \ref{fig1}(c) for unstable single-color vortices. In
particular, Fig. \ref{fig6}(a) implies that the oscillating mode keeps its
vorticity.

\begin{figure}[tbp]
\begin{center}
\includegraphics[height=4.5cm]{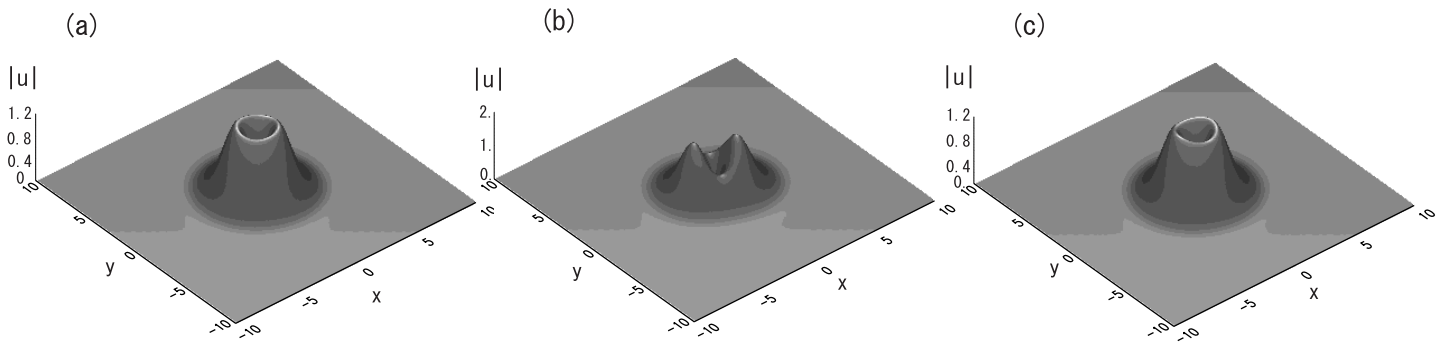}
\end{center}
\caption{The regime of periodic splittings and recoveries of an unstable
two-color vortex (the same one whose evolution is presented in Fig. \protect
\ref{fig6}) is illustrated by a sequence of profiles of $\left\vert u\left(
x,y,z\right) \right\vert $ at $z=90$ (a), $140$ (b), and $190$ (c).}
\label{fig7}
\end{figure}

In fact, the oscillating vortices undergo periodic splitting and recoveries.
An example of this generic dynamical regime as shown in Fig. \ref{fig7} for
the same mode whose evolution is presented in Fig. \ref{fig6}. A similar
persistent regime was found in the 2D model with the self-attractive cubic
nonlinearity, above the threshold of the instability of trapped vortices
(see details in Ref. \cite{cubic-in-trap5}). However, in the case of the
cubic equation the splitting-recovery scenario is replaced, at still larger $%
N$, by the onset of the collapse, which does not occur in the $\chi ^{(2)}$
system.

\section{Conclusion}

The objective of this work is to demonstrate the possibility of the
stabilization of $\chi ^{(2)}$ solitary-vortex modes by the isotropic
trapping potential. The existence and stability of various modes supported
by the system is determined by the total power, $N$, mismatch $q$, and modal
topological charge, $m/2$ ($m$ is the intrinsic vorticity of the SH
component). Using, in parallel, numerical solutions and static and dynamical
versions of the VA (variational approximation), we have found that, at $%
N<N_{c}^{(m)}(q)$, all modes are of the single-color type, represented by
the SH (second-harmonic) component only. In contrast with the usual
assumption that such modes are subject to the parametric instability against
the generation of the FF (fundamental-frequency) field, we have
demonstrated\ that they are \emph{stable} at $N<N_{c}^{(m)}(q)$, including
the (formal) \textit{semi-vortex} with topological charge $1/2$ ($m=1$).
However, $N_{c}^{(m)}(q)\equiv 0$ at $q>1$, i.e., the single-color modes are
indeed completely unstable at large $q$. The modes with $m=0$ and $m=2$
undergo pitchfork bifurcations exactly at $N=N_{c}^{(m)}$, which,
destabilizing the single-color states, give rise to stable two-color
complexes. For $m=0$, the emerging states are always stable, while the
vortical two-color mode, with $m=2$, has an upper stability limit, $%
N=N_{c2}(q)$. At $N>N_{c2}(q)$, the unstable vortices feature
periodic splittings and recoveries, keeping their topological
charge. In addition to optics, these results may be realized in
atomic-molecular BEC mixtures. The semi-vortex does not bifurcate at
$N=N_{c}^{(m=1)}$; instead, it develops persistent quasi-chaotic
oscillations, involving additional angular harmonics in both the SH
and FF fields, and features a rotating ``groove", which resembles
the mixed screw-edge dislocation induced by the semi-integer
vorticity.

This work suggests possibilities for the analysis in other directions. The
stability of trapped vortices and semi-vortices with higher values of the
topological charge ($m/2=1.5,~2,~2.5,~3,...$) may be a natural
generalization, as well as the consideration of trapped modes in the
framework of the Type-II (three-wave) $\chi ^{(2)}$ system. On the other
hand, it may be easy to perform a similar analysis for fundamental and
higher-order odd and even (spatially antisymmetric and symmetric,
respectively) modes in the 1D version of the $\chi ^{(2)}$ system. In
particular, the trapping potential may have a chance to stabilize the 1D odd
modes, which are always unstable in the free space \cite{Werner}. On the
other hand, a challenging problem is to find stable solutions for 3D
``bullets" in the extended version of system (\ref{eqs}), including the
temporal variable and group-velocity-dispersion terms. The possibility of
the existence of stable ``bullets" with embedded vorticity is an especially
intriguing issue (cf. Ref. \cite{3Dvortex}, where stable ``spinning bullets"
were found in the free-space model combining the $\chi ^{(2)}$ and
self-defocusing $\chi ^{(3)}$ nonlinearities). The 3D setting makes it also
possible to study collisions between stable solitons moving along the
trapping potential pipe.

\end{document}